# Brightening dark trions in WS$_2$ monolayers via introducing atomic sulfur vacancies

Xuguang Cao[1], Wanggui Ye[1], Debao Zhang[1], Ji Zhou[1], Lei Peng[2], Changcheng Zheng[3], Kenji Watanabe[4], Takashi Taniguchi[5], Jiqiang Ning[1, *], and Shijie Xu[1, *]

[1]Department of Optical Science and Engineering, College of Future Information Technology, Fudan University, 2005 Songhu Road, Shanghai 200438, China

[2]Key Laboratory for Computational Physical Sciences (MOE), Institute of Computational Physical Sciences and Department of Physics, Fudan University, Shanghai 200433, China

[3]Division of Natural and Applied Sciences, Duke Kunshan University, Kunshan 215316, China

[4]Research Center for Electronic and Optical Materials, National Institute for Materials Science, 1-1 Namiki, Tsukuba 305-0044, Japan

[5]Research Center for Materials Nanoarchitectonics, National Institute for Materials Science, 1-1 Namiki, Tsukuba 305-0044, Japan

*Corresponding authors, emails: jqning@fudan.edu.cn; xusj@fudan.edu.cn

## Abstract

Understanding the effects of atomic defects on the optical functionality of two-dimensional (2D) layered materials is critical to develop novel optical and optoelectronic applications of these ultimate materials. Herein, we correlate sulfur vacancies (V$_S$) and luminescence properties of dark trions in monolayer WS$_2$ through introducing V$_S$ defects and conducting a systematic optical spectroscopic characterization at cryogenic and room temperatures. It is unraveled that the V$_S$ defects can brighten the dark trions via introducing a stronger spin-orbit coupling due to the space inversion symmetry broken by the defects. Furthermore, the wavefunction localization of the dark trions bound at V$_S$ defects results in significant enhancement of the phonon scattering from the K$_2$ valley phonons and hence makes the K$_2$ phonon replica dominant in the emission spectrum. Theoretical calculations of the temperature-dependent photoluminescence spectra with quantum mechanics-based multimode Brownian oscillator model show strong support for the above arguments. Brightening the dark excitons not only sheds light on the understanding of the intriguing excitonic properties of 2D semiconductors, but also may open a way for regulating the optoelectronic performance of two-dimensional semiconductors.

Atomically thin transition metal dichalcogenides (TMDCs) are extraordinary two-dimensional (2D) semiconductors with strong exciton effects due to prominent quantum confinement and weak dielectric screening [1-3]. On one hand, the strong exciton effects in TMDC materials make them an excellent platform for studying novel exciton physics, such as the photophysics of excitons, trions (one kinds of charged excitons), and even their coupling composite quasiparticles [4]. On the other hand, the formation of these tightly bound excitons originating from the direct optical bandgap transitions located at two time-reversal K/K' valleys of the hexagonal Brillouin zone, arouses interest in valley electronics for quantum information processing applications [5-7]. Moreover, strong spin-orbit coupling (SOC) induced by the lack of space inversion symmetry splits the band edge electrons with two opposite spin states [8, 9]. For example, the resulting valence (conduction) band splitting energy is about 400 (30) meV at the K/K' points [10]. The splitting sub-bands render two spin configurations of the direct excitons, where the excitons with the same band spin ($S_z = 0$) are optically bright, whereas the excitons with the opposite band spin ($S_z = \pm 1$) are optically dark for the electron spins polarized along the out-of-plane ($z$) direction [8]. The selection rules based on electric dipole approximation strongly suppress the radiative recombination of the spin mismatch dark excitons, leading to long lifetimes and valley-decoherence times [11, 12]. Recently, a series of techniques, like the two-photon absorption spectrum [13], in-plane (IP) magnetic field brightening method [8], near-field coupling to surface plasmon polaritons [14], and antenna-tip technique based on the Purcell effect [15], have been proposed to enhance the interactions between the optical field and dark excitons such that the additional excitonic information can be accessed. Available extensive exploration has shown that the spin-forbidden dark excitons couple to light solely through a residual out-of-plane (OP) spin-flip dipole matrix element. That is, they exhibit the OP transition dipole moment and the predominant IP emission [16-20]. In particular, the spin-forbidden dark excitons in the tungsten-based TMDC monolayers (MLs) are the energetically low-lying states compared to the bright excitonic states, and a large population may accumulate for them at low temperatures, eventually leading to considerable IP emission, although they possess extremely weak oscillator strength [8, 16, 17, 21]. Furthermore, the trion state was proposed to describe the interactions between the optically excited bright/dark excitons and a Fermi sea of electrons or holes in finite charge density [22-24]. Meanwhile, the dark excitons and trions can achieve spin and intervalley relaxation with the assistance of efficient phonon scattering. Involved phonons include the zone edge phonons and valley phonons describing the collective lattice oscillations at K/K' points [25, 26]. The relaxation process leads to a series of observable phonon replica emission peaks of the dark states, which is confirmed by magneto-optical photoluminescence (PL) spectroscopic measurements [26, 27]. Therefore, exploring the interactions between the valley phonons with large momenta and the dark excitonic states provides a novel manipulation of excitonic relaxation behavior in TMDCs materials.

Additionally, an ultra-thin configuration of the TMDC atomic layers renders the sensitivity behavior of their optoelectronic properties to defects, particularly in various phenomena associated with chalcogen vacancies, such as defect-activated emission [28, 29], single-photon emission [30], and hopping transport via localized gap defect states [31]. Regarding the formation of vacancy defects, the ambient oxidation method recently reported can regulate the generation of chalcogen vacancy defects by the removal of sulfur atoms from S-based TMDC atomic planes [32, 33]. Actually, oxygen molecules in the atmospheric environment can passivate the S vacancy ($V_S$) state due to energetically favorable chemisorption, forming a benign isoelectronic defect [34, 35]. Very recently, a few reports have considered the coupling of dark trions and defects and claimed that defect-assisted recombination may brighten the intervalley dark trions (momentum-forbidden states) [36, 37]. These studies greatly spark interest in exploring the physical mechanisms underlying the interactions between the dark trions and the relevant phonons at $V_S$ defects. Accordingly, the controllable generation and manipulation of the $V_S$ defects may offer sophisticated control for the relaxation and recombination of the dark trions and their phonon replica transitions, and thus provide an excellent chance to get a comprehensive understanding of many-body physics in TMDC materials.

In this article, O-substituted chemisorption on $WS_2$ MLs was introduced by the ambient oxidation method [32], and then the $V_S$ defects were controllably created by removing the O chemisorption on $V_S$ sites through high-intensity laser annealing. Effects of the $V_S$ defects on the luminescence properties of the dark trions and their phonon replicas in the $WS_2$ MLs were systematically investigated through conducting abundant micro-spectroscopic characterization at cryogenic and room temperatures. It is found that the created $V_S$ defects can significantly darken the emissions of bright excitons and their complexes, at the same time can strongly brighten the dark trions and their $K_2$ valley phonon replicas, i.e., by up to two orders of magnitude, such that the phonon replicas become observable even at room temperature (RT). The physical mechanism behind this phenomenon is that the $V_S$ defects break the mirror symmetry and the three-fold rotation symmetry of the ML $WS_2$, thereby introducing a strong IP spin-orbit coupling which mixes the bright and dark excitons by spin-flip processes and hence brightens the dark trion states. In addition, the binding of the dark trions at $V_S$ defects is revealed to greatly enhance the $K_2$ valley phonon-assisted exciton scattering due to the compression of their wave functions, and eventually results in the emission enhancement of their $K_2$ phonon replicas. Furthermore, the temperature-dependent PL spectra of the $V_S$-rich ML sample were calculated by the quantum mechanics-based multimode Brownian oscillator (MBO) model, showing a reduced exciton-phonon coupling strength ( $S$ parameter) with increasing the temperature. The temperature dependence of the PL peak positions of the dark trions is found to be weaker than that of the bright excitons, which may be due to the OP transition dipole moment of the dark trions. These findings shed some light on the complicated many-body interactions between the dark trions and the point defects in two-dimensional TMDC semiconductors and pave the way to regulate the optoelectronic properties of these functional ultimate materials.

Figure 1a presents a schematically atomistic representation of the $V_S$-free ML $WS_2$. The bright exciton and trion are schematically drawn in the ML $WS_2$ plane. As seen in Figure 1b, a dark trion (or its $K_2$ phonon replica) schematically binds at a $V_S$ defect in $V_S$-rich ML $WS_2$. To investigate the effects of $V_S$ defects on the excitonic properties of the ML $WS_2$, the optical spectra of the pristine and aged $WS_2$ samples were systematically measured and compared. The pristine ML $WS_2$ was prepared by immediately transferring the MLs exfoliated from the bulk crystal $WS_2$ onto a $SiO_2/Si$ substrate, while the aged ML WS$_2$ was obtained by oxidating the prepared pristine WS$_2$ in the atmosphere for about 2 months, accompanied by the process of O-substituted chemisorption with the removal of sulfur atoms [32].

Figure 1c shows a schematic illustration of the neutral exciton (A), singlet trion (T$_s$), triplet trion (T$_t$), and charged biexciton (AT). For the A excitons, they exhibit optically allowed transitions, generating a dominant peak in the excitation-power-dependent PL spectra of the pristine ML WS$_2$ at RT, as sketched in the left panel in Figure 1d. As the excitation power was increased from 1 to 450 μW, a weak structure at 32 meV below the A exciton peak was observed, which is assigned to the emission of bright trions including the singlet intravalley trion T$_s$ and the triplet intervalley trion T$_t$, as schematically illustrated in Figure 1c [38-40]. For comparison, the PL spectra of the aged ML WS$_2$ were depicted in the right panel in Figure 1d for different excitation powers. Under the weak excitation (i.e., 1 μW) and other identical conditions, the PL spectrum of the aged ML WS$_2$ is also dominated by the A exciton emission. However, the spectral characteristic of the aged ML exhibits a weaker signal intensity and broader linewidth compared to those of the pristine one, which is consistent with the optical response of the oxidized WS$_2$ reported previously [32, 41]. Measured X-ray photoelectron spectroscopy (XPS) shows the presence of W-O bonds and oxygen molecules in the aged WS$_2$ layers, indicating the occurrence of O-substituted chemisorption and oxygen physisorption, as seen in Supplementary Figure S1. Unlike the pristine ML, the increase of excitation intensity renders a succeeding change of the dominance peak from the A exciton emission to the low-energy side signal (T$_s$/T$_t$) in the aged ML WS$_2$. Under the medium excitation, the T$_s$/T$_t$ peak is dominant over the A peak. Eventually, the dominance peak becomes the T$_{K_2}^D$ peak below 68 meV of the A emission under the high-power excitation (i.e., 450 μW). It should be noted that a blueshift trend of the A exciton peak seems to happen in the aged ML, as indicated by a purple arrow in the right panel of Figure 1d. Therefore, the direct influence of the focus laser heating effect can be excluded in the interested excitation power range in atmosphere conditions. However, local laser heating, under a vacuum condition, has confirmed to be an indeed effective annealing method to realize the removal of the physical and chemical absorption of atmospheric molecules, thereby indirectly affecting the spectral response of the ML WS$_2$ [42-44]. To confirm the potential mechanism, the PL spectra of the aged ML sample were measured at different vacuum levels. It is found that when ambient air was expelled, the gas desorption on ML WS$_2$ occurred, accompanied by the rapid quenching of the A exciton emission intensity [45]. Under low air pressure (i.e., 1.5 torr), the emission intensity of the low-lying peak, comprising the bright trion and AT (with binging energy of 50 meV), was comparable to that of the A excitons, as shown in Supplementary Figure S2a. Correspondingly, through irradiating the aged ML WS$_2$ and monitoring its PL spectra in vacuum conditions, rapid decline of the A exciton peak can also be observed, as the result of the gas desorption due to laser annealing (Supplementary Figure S2b and S2c). Meanwhile, the rapidly reduced PL weight of the A excitons is consistent with the changes in spectral characteristics caused by the introduction of V$_S$ defects due to oxygen desorption, as previously reported [29, 46, 47]. Therefore, in the aged ML WS$_2$, the high-intensity laser irradiation renders the effective dissociation of O-chemisorption at V$_S$ defects, as well as the exposure of the V$_S$ defects in the atomic interface, being similar to the effect of high vacuum annealing at high temperatures [35]. For the aged samples after laser treatment in a vacuum, regardless of the substrate configuration, i.e., suspended or SiO$_2$/Si substrate-supporting ML WS$_2$, their PL spectra measured under weak excitation intensity always exhibit the dominant $T_{K_2}^D$ emission, as shown in Figure 1e. More importantly, after the ML sample with laser annealing was removed from the vacuum and exposed to an oxygen-containing atmosphere, already-disappeared emissions of the A excitons and trions can slowly recover and gradually become dominate in the PL spectrum (Supplementary Figure S3), indicating the reversible manipulation of the luminescence behavior of the excitons by passivating and introducing V$_S$ defects.

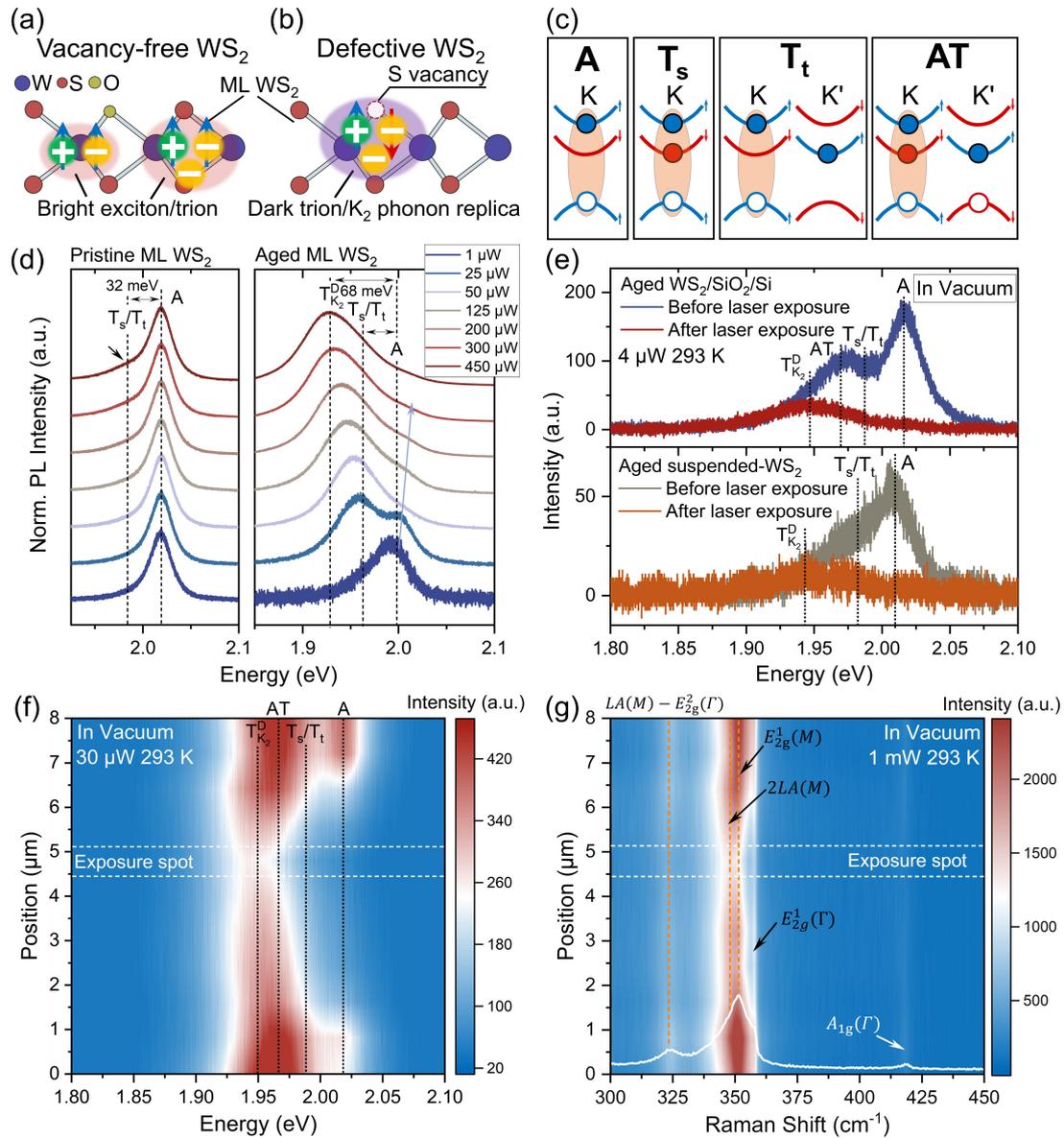

**Figure 1. (a)** Schematic diagram of bright exciton and trion in $V_S$ defect-free ML $WS_2$. **(b)** Schematic diagram of dark trion or its $K_2$ phonon replica in defective ML $WS_2$ with $V_S$ defect. **(c)** Schematic illustration of the neutral exciton (A), singlet trion ($T_s$), triplet trion ($T_t$), and charged biexciton (AT). **(d)** Excitation intensity-dependent RT PL spectra of the pristine (left panel) and aged (right panel) ML $WS_2$ for change of the laser power from 1 to 450 μW. For clarity, the PL spectra are normalized and shifted vertically. **(e)** RT PL spectra of the aged $SiO_2$-supported $WS_2$ (upper panel) and the aged suspended $WS_2$ (lower panel) before and after high-intensity laser exposure, measured under the excitation of a 532 nm laser with power of 4 μW. **(f)** Spatially dependent RT PL spectra measured on the aged ML $WS_2$ after high-intensity laser

exposure in vacuum conditions under the excitation of 532 nm laser with power of 30 µW. **(g)** Spatially dependent RT Raman spectra measured with the excitation laser power of 1 mW on the aged ML $WS_2$ after high-intensity laser exposure in vacuum conditions. The laser exposure spot is within the region outlined by two horizontal dashed lines. The white solid line is a representative Raman spectrum.

Under vacuum conditions, spatially dependent RT PL spectra were measured via line-mapping under the 30 µW laser excitation on the aged ML sample after laser annealing, as sketched in Figure 1f. Within a spatial range of about 1 µm of the laser exposure spot, the PL spectra with relatively lower intensity are dominated by the $T_{K_2}^D$ emission peak, suggesting that the focus laser heating be used as an effective annealing method for regulating the optoelectronic properties of TDMCs and even for fabricating TMDCs-based devices. Clear Raman signals can still be observed, as shown in the space- and time-dependent Raman spectra (Figure 1g and Supplementary Figure S1d), indicating that the CW laser with the relatively low energy density, usually used in the spectroscopy measurements, generally does not damage lattice structure of ML semiconductors unlike femtosecond intense laser [48]. Furthermore, the surfaces of the pristine MLs exhibit more thermodynamically stable behavior than those of the defective counterparts. $V_S$ defects tend to significantly increase the surface reactivity and the oxygen adsorption and desorption rates at vacancies [33], making the PL spectral characteristic of the aged $WS_2$ more sensitive to the laser annealing (Figure 1d).

To correlate the $V_S$ defects and the emergence of the low-lying $T_{K_2}^D$ emission peak, we performed a systematic spectral measurement on the studied samples at cryogenic temperature. Figure 2a shows the cryogenic PL spectrum of the pristine ML $WS_2$ measured with weak excitation intensity (1 µW). The spectrum can be deconvoluted by multiple Gaussian line-shape functions, comprising multiple peaks of the bright A excitons, $T_s/T_t$, AT excitons, and several dark trions. Among them, the PL peaks of the A excitons and the $T_s/T_t$ excitons have corresponding resonance structures in the cryogenic reflectance spectrum [49]. The dominant AT emission feature with a binding energy of 50 meV is well consistent with the spectral structure in previouly reported literature [50]. $T_D$ peak can be further confirmed by the magneto-PL spectrum of the sample via applying a parallel magnetic field [17], as shown in Supplementary Figure S4. The cryogenic PL spectrum of the aged $WS_2$ exhibits almost the identical spectral characteristics as the pristine ML sample. Consistently with the RT spectra, the cryogenic PL spectra of the aged $WS_2$ ML also evidence the passivation effect of oxygen chemisorbed on the $V_S$ sites [35]. Colorful maps of the laser-power-dependent PL spectra of the pristine and aged ML $WS_2$ are illustrated in Figures 2c and 2d, respectively. Similarly to the RT PL measurements, as the laser power increases, intensities of all the cryogenic PL peaks of the pristine $WS_2$ proportionally increase. However, in the aged $WS_2$, the dominant peak gradually changes from the AT emission to the low-lying peaks with increasing excitation power. After exposure to a high-intensity laser (400 μW), the characteristics of the PL and reflectance spectra of the pristine $WS_2$ show no significant changes (maroon curves in Figure 2a). But the reflectance and PL spectra of the aged $WS_2$ after laser annealing show significant changes (brown curves in Figure 2b). For example, the A exciton resonance structure in the reflectance spectra exhibits a 9 meV blueshift and a decreased amplitude. The blueshift is also observed for the A exciton PL peak in the varying-power PL spectra at RT in the aged $WS_2$. This phenomenon may be explained as the increased $V_S$ density, that is, the decreased ML mass density, reduces the dielectric screening suffered by the hydrogenic excitons and the excitonic oscillator strength [2, 51]. Correspondingly, the bright exciton peaks, including the A excitons, $T_s/T_t$, and AT biexcitons, almost disappear, and are replaced by the low-lying peaks. Based on the excitation power-dependent PL spectra in Figure 2d, the low-lying peaks are judged to originate from the emission features of the $T_D$ and $T_{K_2}^D$ in the pristine sample. The weak $T_{K_2}^D$ structure with an energy of 19 meV lower than $T_D$ peak corresponds to the previously denoted $T_{K_2}^D$ peak observed in the RT spectra in the aged $WS_2$ after laser annealing.

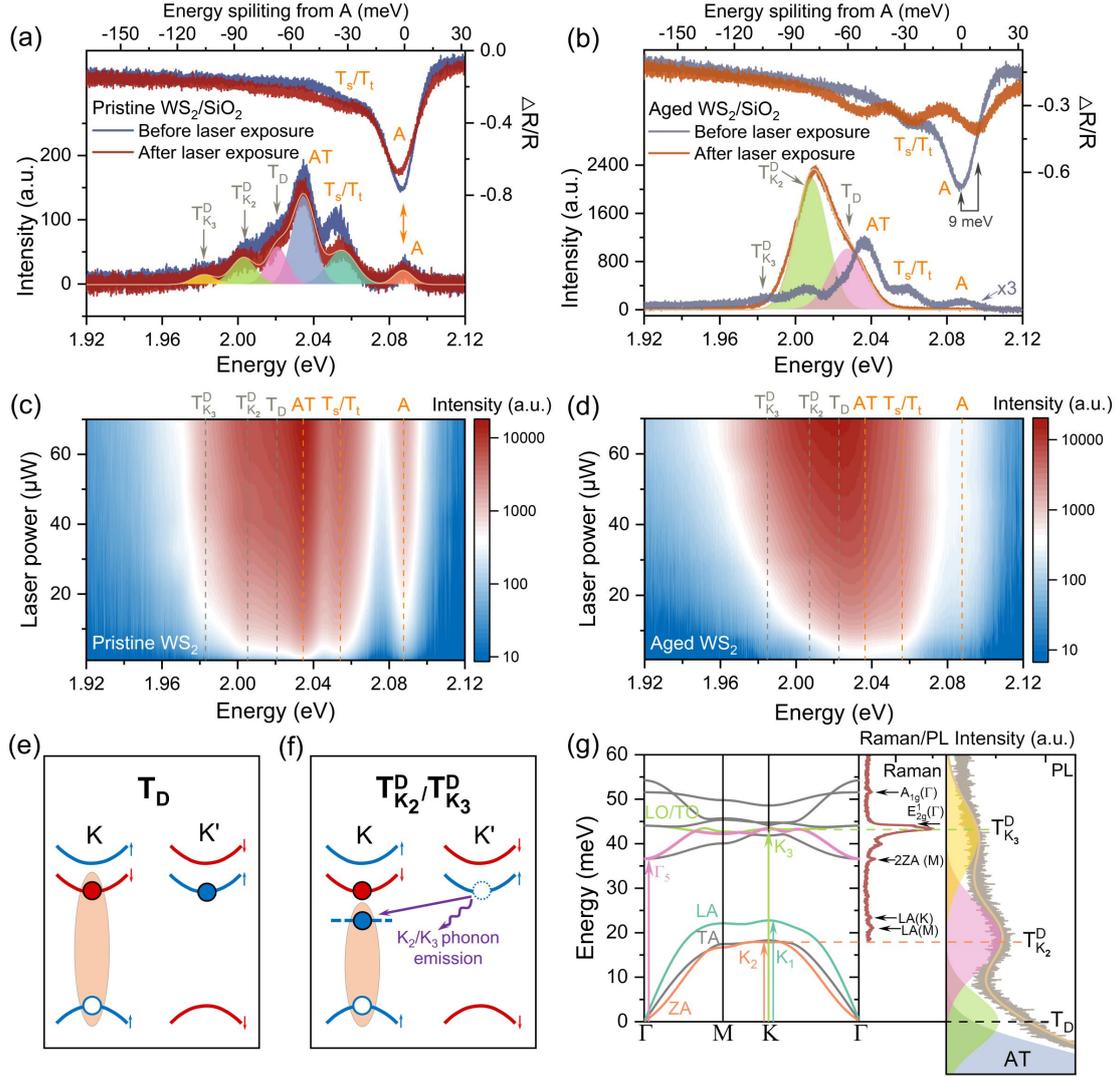

**Figure 2.** Reflectance spectra (upper lines) and PL spectra (lower lines) of the pristine **(a)** and aged **(b)** ML WS$_2$ before (deep blue and gray lines) and after (deep red and brown lines) high-intensity laser exposure. The PL spectra (noisy curves) were measured under the weak excitation intensity of 1 μW at 6 K. Fitting curves (filled colorful smooth curves) and cumulative curve (light yellow smooth curve) with Gaussian line shape function are shown in the figure. Colorful maps of the excitation intensity-dependent PL spectra of the pristine **(c)** and aged **(d)** ML WS$_2$ measured at 6 K with the laser power range from 1 to 70 μW. Schematic illustrations of the T$_D$ trion **(e)** and its T$_{K_2}^D$/T$_{K_3}^D$ phonon replica **(f)**, respectively. The intervalley transition (purple straight arrow) of the K' valley electron (open blue dashed circle) by emitting a K$_2$/K$_3$ valley phonon (purple wave arrow) to a virtual state (blue dashed lines). **(g)**

Calculated phonon dispersions (left panel) and RT Raman spectrum (middle panel) of the ML WS$_2$. Cryogenic PL emission peaks (right panel) of the T$_D$, T$_{K_2}^D$ and T$_{K_3}^D$ phonon replicas with the T$_D$ peak position energy as the reference point.

To further identify the spin and valley origins of the constituent electrons and holes in T$_{K_2}^D$ and T$_{K_3}^D$ emission peaks, magneto-PL measurements are very helpful to extract their effective Landé $g$-factors [52, 53]. Prando et al. reported the OP magnetic field-dependent cryogenic PL measurements of WS$_2$ MLs on the Talc (Mg$_3$Si$_4$O$_{10}$(OH)$_2$) dielectric material, and observed significant Zeeman shifts in peak features during the application of OP magnetic fields up to 30 T [54]. The $g$-factor of the T$_D$ peak was reported as -9.23 [54]. The T$_D$ peak was identified by them as the emission from the intravalley spin-forbidden dark trion with the opposite band spins, as schematically shown in Figure 2e. Meanwhile, the T$_{K_2}^D$ and T$_{K_3}^D$ with $g$-factors of -12.34 and -12.36 are in accord with the proposed theoretical expectation of -13.92 [52-54], assigned to the recombination between electrons and holes from opposite valleys but with the same band spins [26, 52, 54]. Detailed contributions to Zeeman shifts of spin, valley magnetic moment, and atomic orbital magnetic moment are schematically illustrated in Supplementary Figure S5. The large $g$-factor of T$_{K_2}^D$ indicates that the emission feature is extremely possible to derive from the T$_D$ exciton complexes, especially the phonon replicas of the T$_D$. To confirm this speculation, as presented in Figure 2g, various phonon modes in the calculated phonon dispersions by the first principles calculations correspond well to the phonon modes observed in the experimental Raman spectrum in the middle panel, such as the LA (M), LA (K), 2ZA (M), E$_{2g}^1$ ($\Gamma$), and A$_{1g}$ ($\Gamma$). In the cryogenic PL spectrum with the T$_D$ emission peak energy as the reference point, the position energies of the T$_{K_2}^D$ and T$_{K_3}^D$ peaks are lower by 19.0 and 41.0 meV, respectively, which correspond well to the K$_2$ (ZA (K)) and K$_3$ (E"(K)) valley phonon energy of 17.9 and 43.4 meV [55]. It has been shown that electrons can achieve the spin-conserving intervalley scattering assisted by emitting the K$_2$ and K$_3$ valley phonons [25]. As shown in Figure 2f, in the initial configuration of the T$_D$ trions, electrons (open blue dashed circle, located in the opposite valley to holes) with the same spin as holes are scattered into a virtual state of another valley (purple straight arrow), and then radiatively recombine with holes to emit the $T_{K_2}^D$ and $T_{K_3}^D$ phonon replicas (purple wavy arrow) with the lower energies of 19.0 and 41.0 meV, respectively.

Recently, it has been well demonstrated that a hexagonal boron nitride (hBN) layer can provide an atomically flat interface in the hBN-encapsulated ML WS$_2$ showing narrow PL lines. Therefore, a sandwich sample of hBN/WS$_2$/hBN was prepared, and its cryogenic PL and reflectance spectra were measured and depicted in the top figure in Figure 3a. Meanwhile, its schematic atomistic configuration is drawn in the top figure in Figure 3b, correspondingly. In the cryogenic PL spectrum (blue line), the emission peaks of the A exciton the split $T_s$ and $T_t$ due to the exchange interaction [38, 39], AT, $T_D$, $T_{K_2}^D$ and $T_{K_3}^D$ etc., as well as the additional fine structures of the neutral dark exciton ($A_D$), $T_{K_1}^D$ and $T_{\Gamma_5}^D$ are clearly observed. The A exciton resonance structure in the reflectance spectrum (yellow line) can be also well resolved. The $T_D$ emission energy is set as the reference 0 meV for better comparison, as shown in the upper coordinate of Figure 3a. The $T_{K_2}^D$, $T_{K_1}^D$, $T_{K_3}^D$ and $T_{\Gamma_5}^D$ valley phonon replicas are located at 19.0, 23.4, 37.3, and 43.8 meV below the $T_D$ emission peak, respectively. They are ascribed to the phonon replicas of the dark trions $T_D$ by emitting $K_2$ (ZA(K), 17.8 meV), $K_1$ (LA(K), 22.7 meV), $\Gamma_5$ (E''(Γ), 36.6 meV) and $K_3$ (E''(K), 43.4 meV) valley phonons, respectively [26, 27, 56]. Calculated dispersions of these phonon modes can be found in Figure 2g. For the aged ML WS$_2$ sample, the O-substituted chemisorption occurs on the ML surface, as schematically indicated by the red dashed rectangle in the second panel of Figure 3b. The $T_D$ peak deconvoluted by multiple Gaussian line-shape functions is aligned with that of the hBN/WS$_2$/hBN sample. Meanwhile, its $T_{K_2}^D$ and $T_{K_3}^D$ phonon replicas are also well in alignment with those of the hBN/WS$_2$/hBN sample. The third panel of Figure 3b represents the atomistic diagram of the dissociation of the O-substituted chemisorption caused by the high-intensity laser treatment in the aged ML WS$_2$, as previously mentioned. And the $T_D$ and $T_{K_2}^D$ emission peaks dominate the PL spectrum, and their intensities increase by two orders of magnitude compared with those of the aged sample before laser annealing. At the same time, the PL peak of the high-energy bright excitons becomes unobservable. When the aged ML sample was exposed to a dry oxygen atmosphere at RT, however, due to the oxygen chemisorption at the $V_S$ sites, the emission peaks of the bright excitons and complexes reappeared in the cryogenic PL spectrum. Correspondingly, the $T_D$ and $T_{K_2}^D$ emission peaks decrease to their original intensities, as seen in the fourth panel of Figure 3a. Note that due to the dielectric screening difference in the samples, the binding energies of the trion, AT, and $T_D$ exhibit some slight differences. Definitely, the energy separations between the identified phonon replicas and $T_D$ state in these samples are almost identical, as a result of the insensitivity of phonon mode energies to the dielectric screening. These exciting phenomena indicate that the $V_S$ defects exert a profound impact on dark excitons and their complexes. Physical mechanism of the $V_S$ defects mediated the emission enhancement of the dark trions is that the formation of the $V_S$ defects breaks the space inversion symmetry of the ML $WS_2$, and further introduces a stronger spin-orbit coupling [57, 58]. As a result, the conduction band minimum at K/K' valley will further split the spin states, which will enhance the spin-mixing of dark and bright excitons by spin-flip processes, thereby brightening the dark states, especially at low temperatures [8, 59]. Typically, the selection rules demonstrate that only the $K_3$ valley phonon mode, not the $K_2$ phonon, can involve the intervalley electron transitions with spin-conserving between the conduction band edges of high symmetry K/K' points [25, 26]. But the strong dominance of the $T_{K_2}^D$ phonon replica indicates the breaking space inversion symmetry and the localization of dark trions at $V_S$ defects, which could amplify the intervalley electron transitions by the $K_2$ valley phonons [26]. The Bohr radius of the localized dark trions will reduce. Therefore, their wave functions in $k$-space will extend, leading to stronger exciton-$K_2$ phonon interactions.

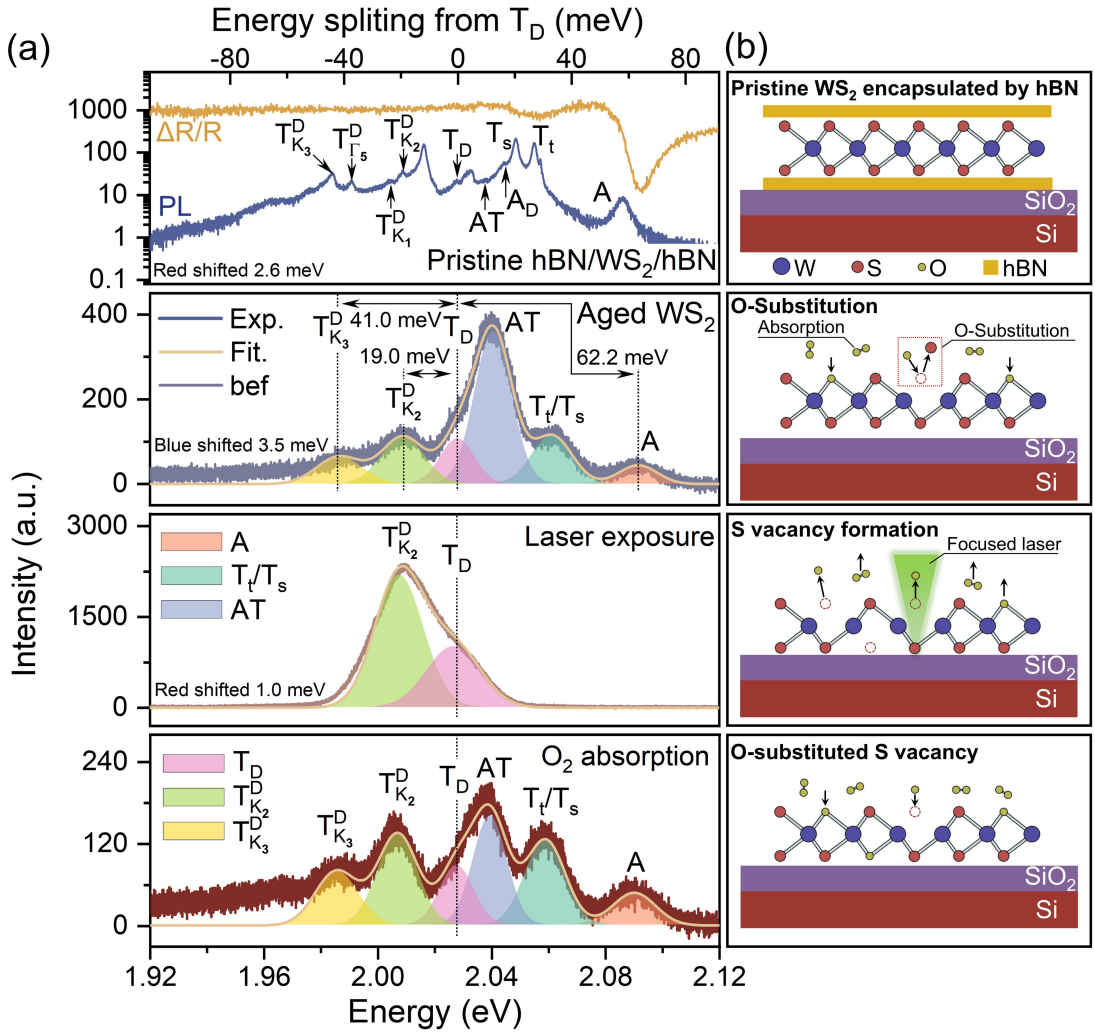

**Figure 3.** Cryogenic PL spectra **(a)** and schematic illustrations **(b)** of the pristine hBN/WS$_2$/hBN, aged ML WS$_2$, aged ML WS$_2$ after laser exposure, and laser-treated ML WS$_2$ with O chemisorption from top to bottom. The T$_D$ emission peak is taken as the reference 0 meV, as shown in the top horizontal axis.

Interestingly, the energy distance between the A excitons and T$_{K_2}^D$ peaks is ∼ 68 meV in the RT PL spectrum, whereas it is increased to be ∼ 80 meV in the cryogenic PL spectrum. This indicates that there may exist a distinct temperature dependence of the transition energies between the bright and dark exciton states. For the aged ML WS$_2$ sample, its temperature-dependent PL spectra are shown in Figure 4a. The A exciton peak pointed by the purple arrows in Figure 4a exhibits a noteworthy redshift and gradually becomes a dominate structure with the rise of temperature. The T$_s$/T$_t$ peak indicated by the light gray arrows shows an obvious thermal quenching behavior, accompanied by a substantial redshift of peak position. The temperature-dependent redshift value of the $T_s/T_t$ peak is almost the same as that of the A exciton, both redshifting about 70 meV from 6 to 300 K. This fact may mean that the binding energy of the trions is insensitive to temperature in the studied aged $WS_2$ sample. And the $T_D$ and $T_{K_2}^D$ emission peaks completely quench before 60 K, while the AT peak disappears at 120 K. By deconvoluting the temperature-dependent PL spectra with multiple Gaussian line-shape functions, the normalized integral intensities (solid circles) of the A and $T_s/T_t$ peaks as a function of temperature are shown in the upper panel of Figure 4c. The solid lines are drawn to guide the eye in the figure (in both cases, the maximum intensity is normalized to unity). The A exciton intensity exhibits a negative thermal quenching till 240 K, related to the supply of low-lying dark excitons and the thermal dissociation of low-energy bright exciton complexes [21]. Subsequently, the intensity gradually decreases with further increasing temperature due to the enhancement of the thermally activated non-radiative recombination. The integral intensities of the $T_s/T_t$ peak show a non-monotonic temperature dependence, i.e., increase first and then start to decrease at 100 K. The phenomenon may be explained as the first feeding of thermally dissociated AT biexcitons and later the thermal dissociation of the $T_s/T_t$ trions. Furthermore, the temperature-dependent PL spectral behavior in the pristine ML is highly similar to the aged one, as shown in Supplementary Figure S6.

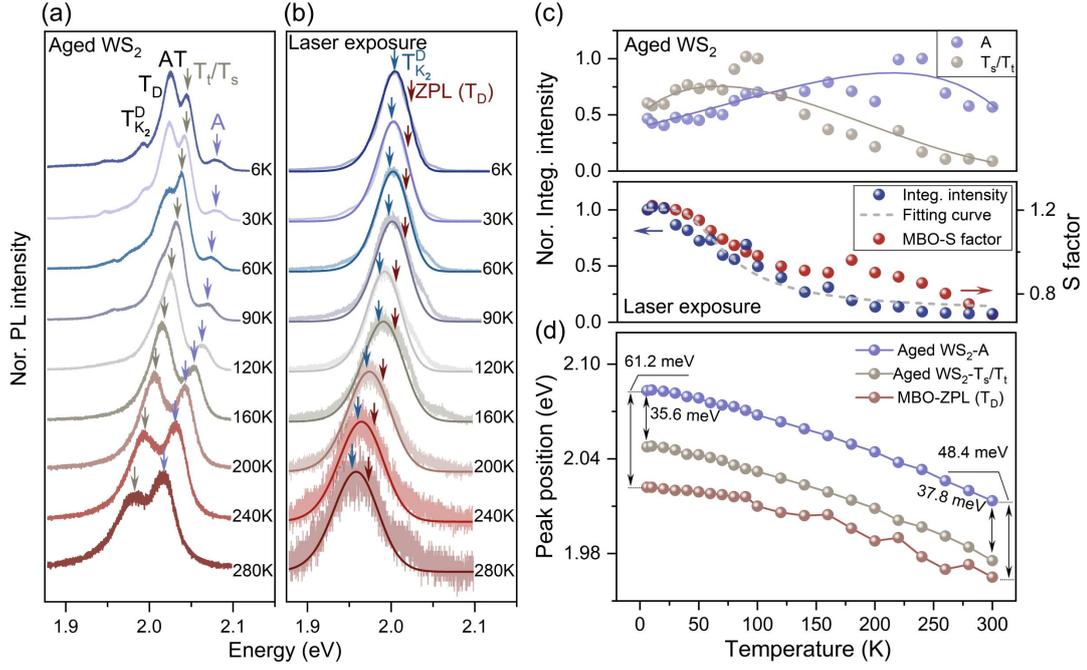

**Figure 4.** Variable-temperature PL spectra measured on the aged ML WS$_2$ with excitation laser power of 4 μW **(a)** and on the aged ML WS$_2$ after laser exposure **(b)**, respectively. **(c)** Normalized integral intensities (solid circles) of the A and T$_s$/T$_t$ excitons in the aged ML WS$_2$ vs. temperature in the upper panel. Solid lines are drawn to guide the eye. Normalized integral intensities (dark blue solid circles) and *S* factor vs. temperature of the laser-treated aged ML WS$_2$ in the lower panel. The light gray dashed line refers to the fitting curve by the formula of $I(T)=I_0/(1+Ae^{(-E_a/k_BT)})$. **(d)** Peak position energies (solid lines with colored circles) of the A, T$_s$/T$_t$, and T$_D$ as a function of temperature.

Shown in Figure 4b are variable-temperature PL spectra (noisy curves) of the aged ML WS$_2$ after laser exposure treatment, composed of the overlapping T$_D$ and T$_{K_2}^D$ peaks within the entire temperature range of interest. The variable-temperature PL spectra exhibit a usual thermal quenching behavior, as represented in dark blue circles in the lower panel of Figure 4c. The thermal quenching behavior may be fitted by the standard formula $I(T)=I_0/(1+Ae^{(-E_a/k_BT)})$. $I_0$ is the intensity at the temperature of 0 K, $E_a$ is the thermal activation energy, $k_B$ is the Boltzmann constant and *T* refers to the temperature. The light gray dotted line in in the lower panel of Figure 4c refers to the fitting curve. The obtained $E_a$ is approximately 18.0 meV, which may represent an average binding energy of the T$_D$ and T$_{K_2}^D$ localized on stationary V$_S$ defects [46]. The center position of the broad peak redshifted only by 46 meV with rising temperature, which is significantly smaller than those of the A and $T_s/T_t$ states in the untreated aged $WS_2$ sample. This could be the origin of the different A- $T_{K_2}^D$ energy separations at low and room temperatures. To quantitatively determine the position energies of the zero-phonon line (ZPL, refers to the $T_D$ emission peak) and extract the dimensionless Huang-Rhys factor (often called $S$ parameter or factor), the multimode Brownian oscillator (MBO) model, considering both the electron-phonon coupling and the phonon-bath mode dissipating effect, was adopted to calculate the theoretical PL spectra. The MBO model was originally proposed by quantum chemists and later employed by Xu et al. to calculate temperature-dependent PL spectra in semiconductors [60, 61], where the calculated PL line shape may be expressed by the following formula:

$$I_{PL}(\omega)=\frac{1}{\pi}\int_0^\infty f(\omega_{eg})\,d\omega_{eg}\text{Re}\int_0^\infty \exp[i(\hbar\omega\text{-}\hbar\omega_{eg}+\lambda)t\text{-}g^*(t)]dt, \qquad (1)$$

where the Gaussian function of $f(\omega_{eg})=\exp[\text{-}2W^{-2}(\omega_{eg}\text{-}\omega_{eg}^0)^2]$ describes the considered inhomogeneous broadening, $\lambda=S\hbar\omega_{ph}$, $\omega_{ph}$ is the angular frequency of the primary phonon, and $\omega_{eg}^0$ is the angular frequency of ZPL, $S$ is the dimensionless Huang-Rhys factor, $g^*(t)$ is the complex conjugate of the spectral response function $g(t)$ that reads

$$g(t)=\frac{1}{2\pi}\int_{-\infty}^\infty d\omega \frac{2\lambda\omega_{ph}^2\omega\gamma}{\omega^2[\omega^2\gamma^2+(\omega_{ph}^2\text{-}\omega^2)^2]}\times[1+\coth(\beta\hbar\omega/2)]\,(e^{-i\omega t}+i\omega t\text{-}1), \qquad (2)$$

where $\beta\equiv1/k_BT$, and $\gamma$ is the damping constant accounting for the low-frequency acoustic phonon bath-induced broadening.

Herein, the characteristic frequency of the primary phonon coupled with the $T_D$ state employed in the MBO calculations was fixed at 142.4 cm$^{-1}$, corresponding to an energy of 17.8 meV of the $K_2$ valley phonon mode. The theoretical PL spectra calculated by the MBO model are shown in deeper-colored lines in Figure 4b, showing a quite good agreement with the experimental data. In the figure, the maroon and dark blue arrows indicate the ZPL and $T_{K_2}^D$ peak positions at different temperatures, respectively. The fitting parameter $S$ physically represents the lattice relaxation amount and the average number of primary phonon modes participating in the luminescence process [62]. Interestingly, the $S$ values tend to decline with the rise of temperature, indicating that the dark exciton-$K_2$ phonon coupling tends to decrease with temperature. The reducing tendency of the $S$ parameter with temperature supports the previous conclusion that the $V_S$ defects boost the localization of the dark trions and hence the dark trion-phonon interactions. With the thermalization of the localized dark trions, their Bohr radius will expand in real space, hence the exciton-phonon coupling weakens, i.e., the $S$ value reduces [63]. Figure 4d describes the temperature dependence of the position energies (solid lines with colored circles) of the A, $T_s/T_t$, and ZPL ($T_D$) peaks. Among them, the A, $T_s/T_t$ positions were obtained from the PL spectra in Figure 4a, while the ZPL positions were extracted from the MBO theoretical calculations. The binding energy of the $T_s/T_t$ excitons is about 35.6 meV at 6 K and insensitive on temperature. Meanwhile, the A-$T_D$ energy distance is approximately 61.2 meV at low temperatures, being consistent with the value obtained in Figure 2b. Furthermore, the A-$T_D$ energy separation significantly reduces with increasing temperature, i.e., decreases by 3.4 meV at 100 K, which is identical to the results of the variable-temperature PL measurements under a high IP magnetic field of 30 T by Kipczak and co-author [64]. The A-$T_D$ energy splitting reaches 48.4 meV at RT, and the obtained peak positions of the corresponding $T_D$ and $T_{K_2}^D$ emissions are 1.965 and 1.947 eV, respectively. The physical mechanism resulting in the temperature evolution difference of the bright and the dark excitons is still unclear, possibly due to their different interactions with lattice phonons [16, 17].

In summary, we systematically investigated the influence of the atomic $V_S$ defects on the exciton behaviors in the ML $WS_2$ on which the $V_S$ defects were controllably formed through the ambient oxidation and laser annealing methods. It is unraveled that the $V_S$ defects brighten the dark trions and their $K_2$ valley phonon replica by up to two orders of magnitude, such that the latter can even be observed at room temperature. These phenomena can be understood in terms of the breaking of space inversion symmetry and enhancement of spin-orbit coupling induced by the $V_S$ defects. The latter results in the mixing of the bright and dark excitons by spin-flipping relaxation, and hence brightens the dark trion states. The wave function localization of the dark trions bound at $V_S$ defects strengthens the scattering from the $K_2$ valley phonons involved within the intervalley excitonic transitions due to the reduction of the Bohr radius of dark trions, eventually enhancing the emission intensity of the $K_2$ phonon replica. The quantum mechanics-based MBO theoretical calculations demonstrate a declining tendency of the exciton-phonon coupling strength ($S$ parameter) with increasing temperatures, show strong support for the above arguments. Our exploration for the correlation between the $V_S$ defects and the optical properties of the dark trions as well as their phonon replicas may offer a novel approach for manipulating excitonic behavior, especially the dark trions' behaviors, and lead to a progress in understanding of the many-body interactions of excitons, phonons, and defects in TDMCs materials.

**Acknowledgements:** The work was financially supported by the National Natural Science Foundation of China (No. 12074324). K.W. and T.T. acknowledge support from the JSPS KAKENHI (Grant Numbers 21H05233 and 23H02052) and World Premier International Research Center Initiative (WPI), MEXT, Japan.

# Supplementary information

**Experimental Section**

**Sample preparation.** Monolayer $WS_2$ flakes were mechanically exfoliated from a high-quality $WS_2$ bulk crystal (HQ Graphene company) by transparent polydimethylsiloxane (PDMS) stamps (Gel-park). The exfoliated monolayer $WS_2$ flakes were identified and verified by optical microscopy, PL, and Raman spectroscopy in the ambient atmosphere at RT. The suspended $WS_2$ samples were fabricated by transferring the PDMS/$WS_2$ onto the $SiO_2$/Si substrate with pre-patterned 10 μm grooves by the reactive ion etching (RIE) procedure. The pristine ML $WS_2$ samples were fabricated by immediately transferring the ML $WS_2$ flakes exfoliated from bulk crystal onto $SiO_2$/Si substrate. And the aged ML $WS_2$ samples were obtained by removing the surface gas adsorption of the exfoliated MLs on PDMS under a vacuum and then placing them in the atmospheric environment for two months to achieve the oxidation progress, following by transferring the oxidated MLs onto a $SiO_2$/Si substrate. The hBN-encapsulated $WS_2$ samples were prepared by transferring pristine ML $WS_2$ onto target hBN layers and then depositing few-layer hBN flakes onto the ML $WS_2$ by the PDMS dry transfer method. Note that before each transferring step, a thermal heating process at around 363 K for 5 minutes was usually conducted to enhance the adhesion force between the $WS_2$ monolayer and hBN or $SiO_2$/Si interface.

**Measurement.** Micro-PL and Raman measurements were performed on a home-assembled multi-function-integrated micro-spectroscopic system equipped with a monochromator (Horiba iHR 550 with 1200 grooves/mm grating), a charge-coupled device (CCD) detector (Horiba Syncerity 7379 with detection wavelength range of 200 - 1000 nm). The excitation source is the 532 nm continuous-wave laser, which was focused on the sample surface via a ×100 objective lens to form an excitation spot size of about 1 μm. The reflectance spectra were conducted by introducing a broadband picosecond laser (NKT, supercontinuum laser) into the above home-assembled system. During the variable-temperature PL measurements, the samples were mounted on the cold finger of a He-closed cycle cryostat (Montana Instruments S100, varying temperature range of 3.2 - 330 K, background vacuum pressure of $10^{-4}$ torr). Low-temperature micro-magneto PL spectra of the pristine ML WS$_2$ under the IP magnetic field were performed in a cryostat equipped with a superconducting magnet in Faraday geometry (Attodry 2100 cryostat, varying temperature range of 1.7 - 300 K and varying vector magnets of 0 - 9 T). XPS measurements were performed by the K-Alpha X-ray Photoelectron Spectrometer System, Thermo Scientific, with an X-ray spot size of 50-400 μm, at RT. The RT PL measurements with different air pressure were conducted in the vacuum chamber of Montana Instruments S100 connected to a mechanical pump, and dry oxygen gas can be controllably introduced into the vacuum chamber.

**Computational Method.** First-principles calculations were performed within the framework of density functional theory (DFT) using the projector augmented-wave (PAW) method [1], as implemented in the Vienna *ab initio* simulation package (VASP) [2]. The exchange-correlation energy was treated using the Perdew-Burke-Ernzerhof functional revised for solids (PBEsol) within the generalized gradient approximation [3]. A plane-wave energy cutoff of 450 eV was employed. The electronic self-consistency and ionic relaxation were considered converged when the total energy and atomic forces reached thresholds of $10^{-6}$ eV and $10^{-4}$ eV/Å, respectively. Brillouin zone integrations were carried out using a 12×12×1 Monkhorst-Pack k-point mesh. A vacuum region of 20 Å was introduced along the out-of-plane direction to prevent spurious interactions between periodic images. Phonon properties were computed using the finite-displacement method as implemented in the Phonopy package [4]. A 3×3×1 supercell was adopted for the force constant calculations, with a 4×4×1 k-point mesh applied for Brillouin zone sampling.

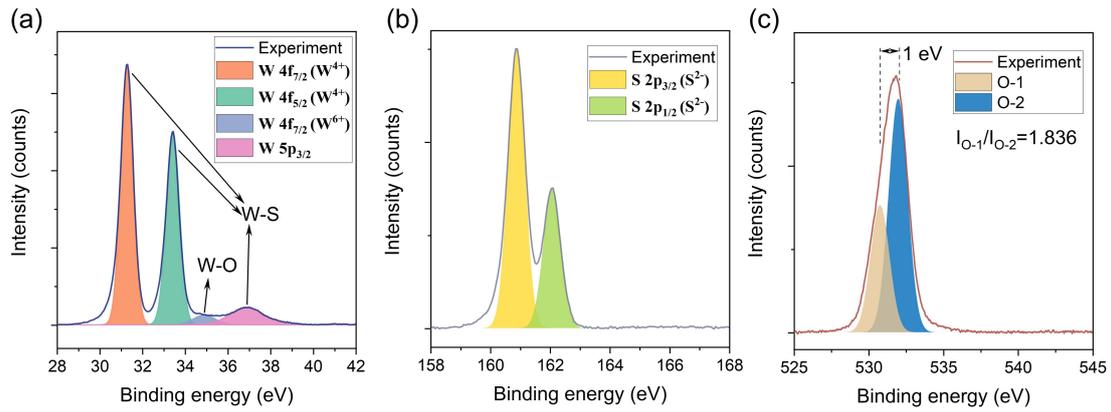

**Supplementary Figure S1.** XPS spectra of the aged WS$_2$ layers showing W 4f and W 5p state [5,6] **(a)**, S 2p state [5,6] **(b)**, and O state [7] **(c)**, with Gaussian-Lorentzian fitting of each peak, exhibiting the possible presence of the W-O bonds and the physisorption of oxygen molecules.

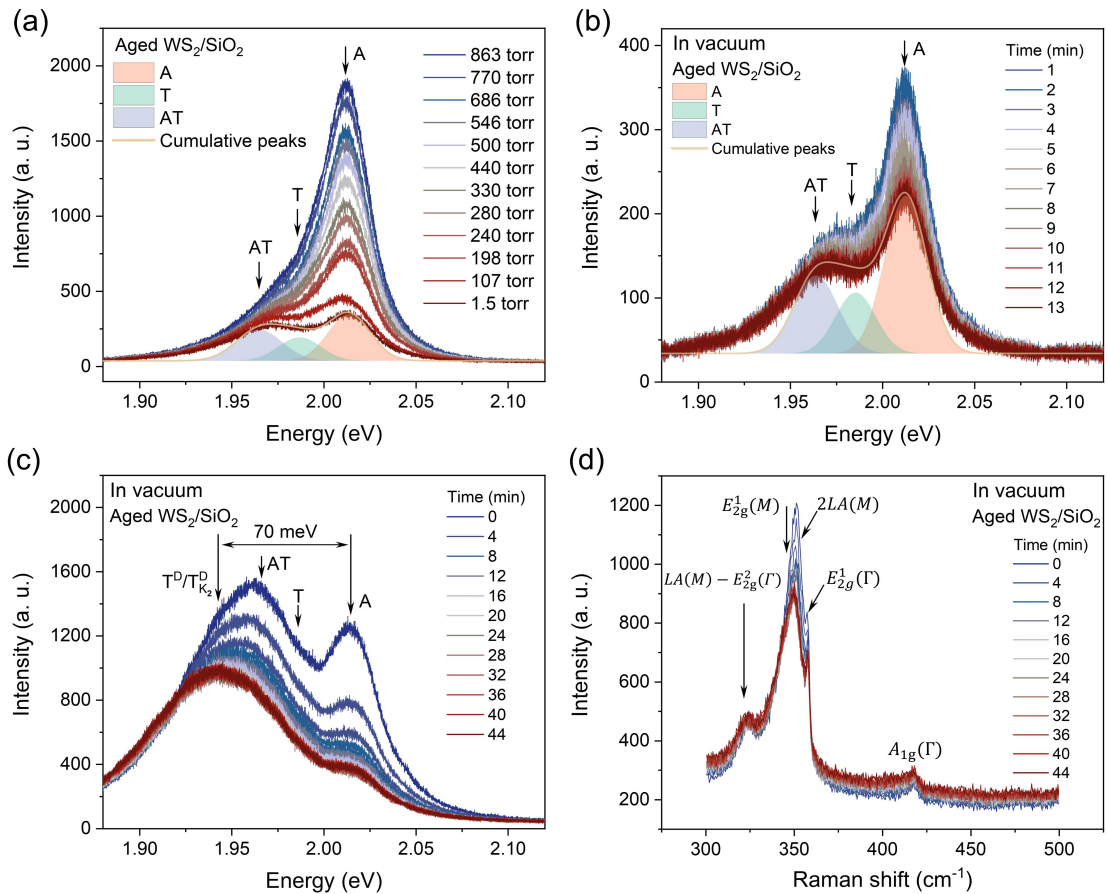

**Supplementary Figure S2. (a)** Air pressure-dependent RT PL spectra of the aged ML WS$_2$ under weak excitation intensity of 4 μW, with fitting curves and cumulative curves (various colored areas) of the PL spectrum measured at 1.5 torr by Gaussian

line shape functions. **(b)** Exposure time evolution of the RT PL spectra in the aged ML WS$_2$ under weak excitation intensity of 4 µW, with fitting curves and cumulative curves (various colored areas) of the PL spectrum with exposure time of 13 minutes by Gaussian line shape functions. **(c)** Exposure time evolution of the RT PL spectra in the aged ML WS$_2$ under high excitation intensity of 400 µW. **(d)** Exposure time evolution of the RT Raman spectra in the aged ML WS$_2$ under weak excitation intensity of 1 mW.

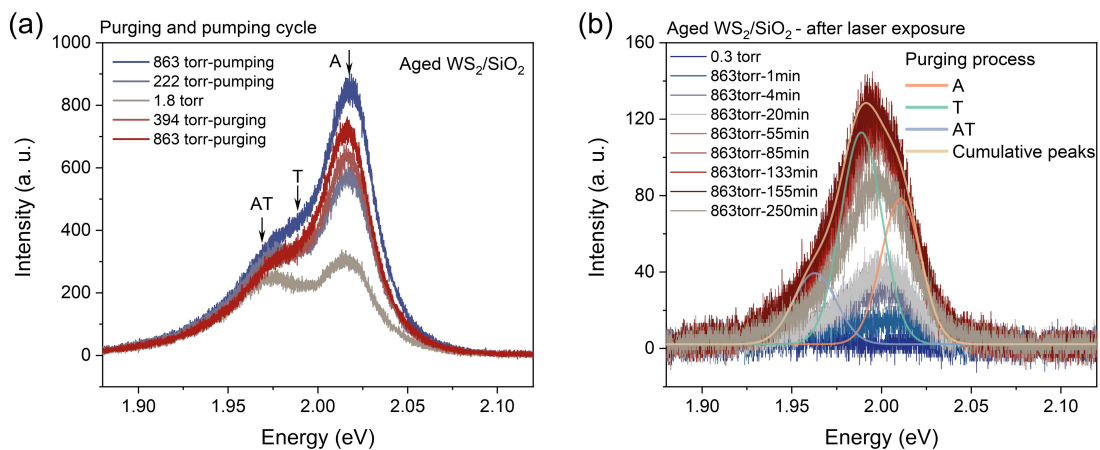

**Supplementary Figure S3. (a)** RT PL spectra of the aged ML WS$_2$ under weak excitation intensity of 4 µW within one purging and pumping cycle. **(b)** Time evolution of the RT PL spectra in the aged ML WS$_2$ with laser annealing after removing vacuum under weak excitation intensity of 4 µW.

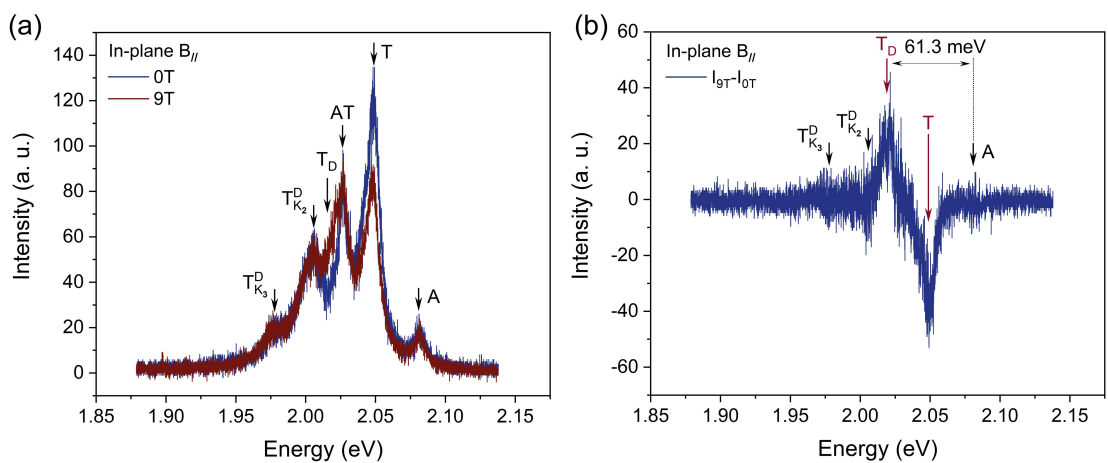

**Supplementary Figure S4. (a)** PL spectra of the pristine ML WS$_2$ under the in-plane magnetic field B$_\parallel$ = 0 T (dark blue line) and B$_\parallel$ = 9 T (maroon line) at a temperature of 2 K with laser power of 4 µW, respectively. The emission intensity of dark trions T$_D$

increases with the increase of the in-plane magnetic field, while bright trions T transform into dark state, resulting in a decrease in emission intensity [8]. **(b)** Corresponding relative intensities defined as $PL_{B=9T} - PL_{B=0T}$, showing the energy distance of 61.3 meV between bright A exciton and $T_D$ dark trions.

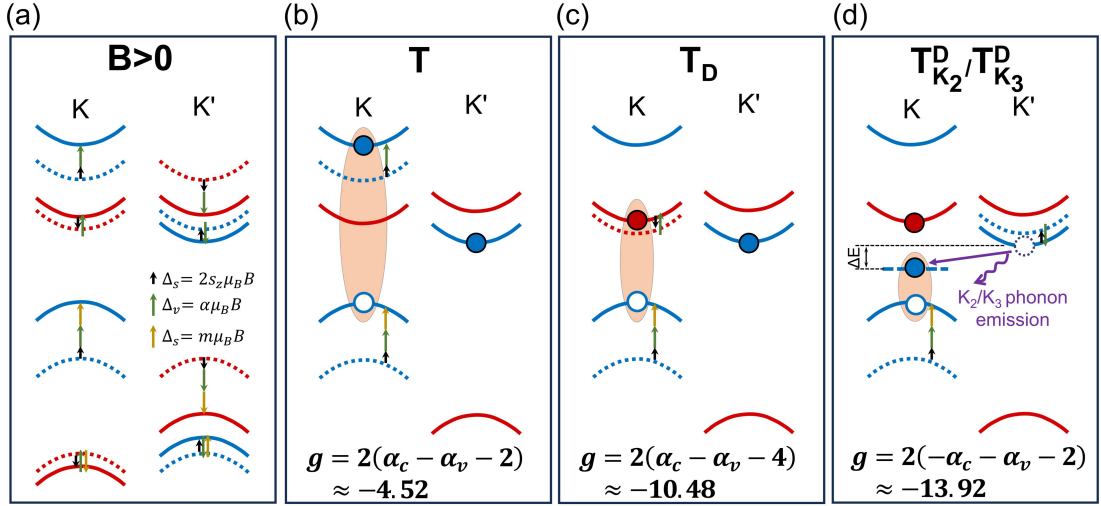

**Supplementary Figure S5. (a)** Band structure at K/K' valley with three main contributions to Zeeman shifts: black arrow for spin magnetic moment, green arrow for valley magnetic moment, and brown arrow for atomic orbital magnetic moment [9]. Different spin-valley configurations of the recombination of T trions **(b)**, TD dark trion **(c)**, and $T_{K_2}^D$ and $T_{K_3}^D$ phonon replicas **(d)**, with corresponding effective Landé g-factors at the bottom.

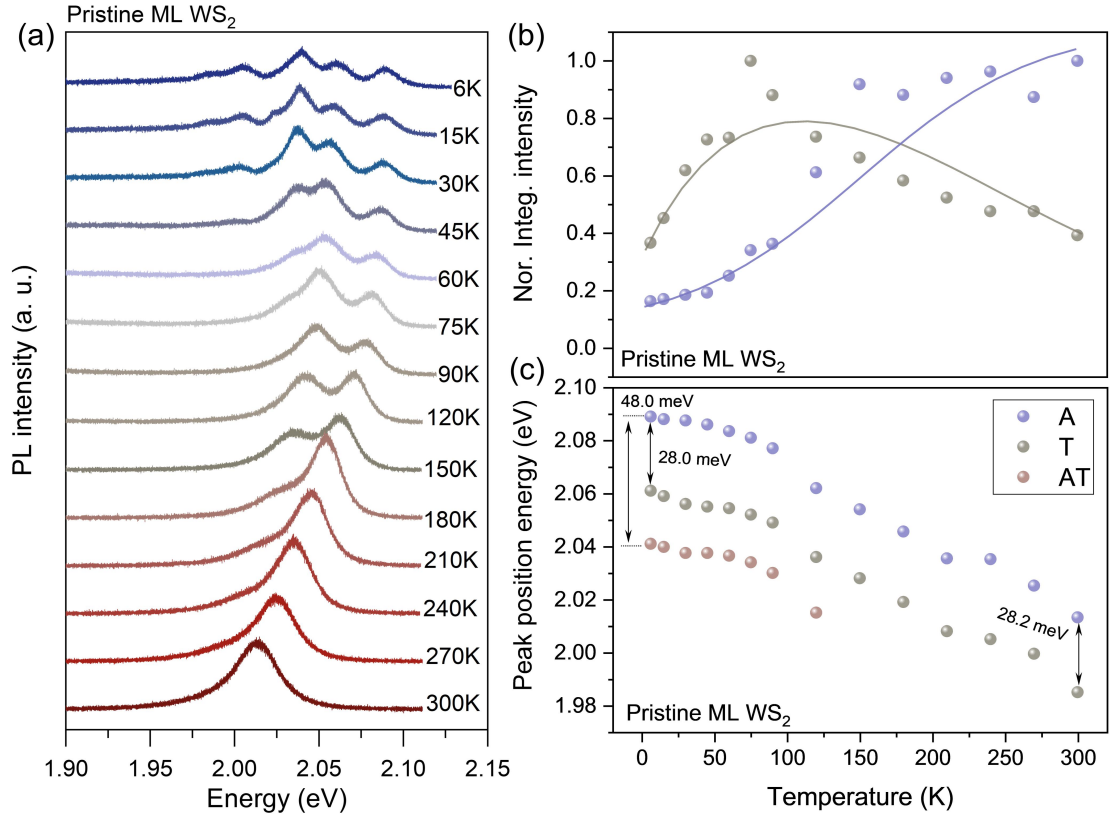

**Supplementary Figure S6. (a)** Temperature dependence PL spectra measured with excitation laser power of 4 μW in the pristine ML WS$_2$. **(b)** Normalized integral intensities vs. temperature of the A, T$_s$/T$_t$ exciton. The solid lines are guides to the eye. **(c)** Peak position energies of the A, T$_s$/T$_t$, and AT exciton as a function of the temperature.